\documentclass[final]{article}
\usepackage{mathrsfs}
\usepackage{amsfonts}
\usepackage{amsmath,amsthm}
\usepackage{amssymb}
\usepackage{amsfonts}
\usepackage[noadjust]{cite}
\usepackage{subfigure}
\usepackage{mathrsfs}
\usepackage{graphicx}
\usepackage{amsmath}
\usepackage{amsfonts}

\newtheorem{remark}{Remark}
\newtheorem{theorem}{Theorem}
\newtheorem{definition}{Definition}
\newtheorem{lemma}{Lemma}

\newcommand{\X}{\mathcal{X}}
\newcommand{\Y}{\mathcal{Y}}
\newcommand{\Z}{\mathcal{Z}}

\renewcommand{\P}{\mathcal{P}}

\newcommand{\V}{\mathcal{V}}

\newcommand{\U}{\mathcal{U}}
\newcommand{\C}{\mathcal{C}}

\begin{document}

%
%
%


\begin{titlepage}
\title{A New Universal Random-Coding Bound for Average Probability Error Exponent for Multiple-Access Channels}
\author{Ali Nazari, Achilleas Anastasopoulos and S. Sandeep Pradhan\\
Electrical Engineering and Computer Science Dept. \\
University of Michigan, Ann Arbor, MI 48109-2122, USA\\
E-mail: \{anazari,anastas,pradhanv\}@umich.edu}
\thispagestyle{empty} \maketitle
\begin{abstract}
 In this work, a new upper bound for average error probability of a two-user
discrete memoryless (DM) multiple-access channel (MAC) is derived.
This bound can be universally obtained for all discrete memoryless
MACs with given input and output alphabets. This is the first bound
of this type that explicitly uses the method of expurgation. It is
shown that the exponent of this bound is greater than or equal to
those of previously known bounds.
\end{abstract}
\end{titlepage}


\section{Introduction} \label{intro}

A crucial problem in network information theory is determining the
average probability of error that can be achieved on a discrete
memoryless multiple-access channel.
More specifically, a two-user DM-MAC is defined by a stochastic
matrix\footnote{We use the following notation throughout this work.
Script capitals $\mathcal{U}$, $\mathcal{X}$, $\mathcal{Y}$,
$\mathcal{Z}$,$\ldots$ denote finite, nonempty sets. To show the
cardinality of a set $\mathcal{X}$, we use $|\mathcal{X}|$. We also
use the letters $P$, $Q$,$\ldots$  for probability distributions on
finite sets, and $U$, $X$, $Y$,$\ldots$ for random variables.}
$W:\X \times \Y \rightarrow \Z$, where the input alphabets, $\X$,
$\Y$, and the output alphabet, $\Z$, are finite sets. The channel
transition probability for sequences of length $n$ is given by
\begin{align}
W^n(\mathbf{z}|\mathbf{x},\mathbf{y}) \triangleq \prod_{i=1}^n
W(z_i|x_i,y_i)
\end{align}
\indent where
\begin{align*}
\mathbf{x}\triangleq (x_1,...,x_n) \in \mathcal{X}^n,
\mathbf{y}\triangleq(y_1,...,y_n) \in \mathcal{Y}^n
\end{align*}
\indent and
\begin{align*}
\mathbf{z}\triangleq (z_1,...,z_n) \in \mathcal{Z}^n.
\end{align*}
It has been proven, by Ahlswede~\cite{Ahlswede71} and
Liao's~\cite{Liao} coding theorem, that for any $(R_X,R_Y)$ in the
interior of a certain set $\mathcal{C}$, and for all sufficiently
large $n$, there exists a multiuser code with an arbitrary small
average probability of error. Conversely, for any $(R_X,R_Y)$
outside of $\mathcal{C}$, the average probability of error is
bounded away from 0. The set $\mathcal{C}$, called \emph{capacity
region} for $W$, is the closure of the set of all rate pairs
$(R_X,R_Y)$ satisfying~\cite{SlWo73}
\begin{subequations}
\begin{align}
0 &\leq R_X \leq I(X \wedge Z|Y,U)\\
0 &\leq R_Y \leq I(Y \wedge Z|X,U)\\
0 &\leq R_X+R_Y \leq I(XY \wedge Z|U),
\end{align} 
\end{subequations}
%
for all choices of joint distributions over the random variables
$U,\ X,\ Y,\ Z$ of the form $p(u)p(x|u)p(y|u)W(z|x,y)$ with $U \in
\U$ and $|\mathcal{U}| \leq 4$. As we can see, this theorem was
presented in an asymptotic nature, i.e., it was proven that the
error probability of the channel code can go to zero as the block
length goes to infinity. Yet, it does not tell us how large the
block length must be in order to achieve a specific error
probability. On the other hand, in practical situations, there are
limitations on the delay of the communication. Additionally, the
block length of the code cannot go to infinity. Therefore, it is
important to study how the probability of error drops as the block
length goes to infinity. A partial answer to this question is
provided by examining the error exponent of the channel.

Error exponents have been meticulously studied for discrete
memoryless channels in point to point data communications. Lower and
upper bounds are known on the error exponent of these channels. A
lower bound, known as the random coding exponent, was developed by
Fano~\cite{FanoBook}. The random coding bound in information theory
provides a well-known upper bound for the probability of decoding
error of the best code, of a given rate and block length. This bound
is constructed by upper-bounding the average error probability over
an ensemble of codes. Gallager~\cite{Gallager-Tightness}
demonstrated that the random coding bound is the true error exponent
for the random code ensemble. This result illustrates that the
weakness of the random coding bound, at low rates, is not due to
upper-bounding the ensemble average. Rather, this weakness is due to
the fact that the best codes perform much better than the average,
especially at low rates. Barg and Forney~\cite{Barg-RandomCode}
investigated two different upper bounds on the average probability
of error, called the typical random coding bound and the expurgated
bound. The typical bound is basically the typical performance of the
ensemble. By this, we mean that almost all random codes exhibit this
performance. In addition, they have shown that the typical random
code performs much better than the average performance over the
random coding ensemble, at least, at low rates. The random coding
exponent may be improved at low rates by a process called
``expurgation'' which yields a new bound that exceeds the random
coding bound at low rates. It has been shown that the expurgated
bound is strictly larger than both the random coding and the typical
random coding bounds at low rates. It has also been demonstrated
that both the expurgated and the typical random coding bounds are
equal at $R=0$. At this specific rate, the upper bound on the
reliability function is also equal to these bounds~\cite[pg.
189]{Csiszarbook}.

In regard to the Multiple-Access Channels, stronger versions of
Ahlswede and Liao's coding theorem, giving exponential upper and
lower bounds for the error probability, have been derived by
numerous other authors. Slepian and Wolf~\cite{SlWo73},
Dyachkov~\cite{Dyachkov}, Gallager~\cite{Gallager-Multiaccess},
Pokorny and Wallmeier~\cite{Pokorney}, and Liu and
Hughes~\cite{Liu-RandomCoding} have all studied upper bounds on the
error probability. Haroutunian~\cite{Haroutunian} and
Nazari~\cite{nazari08} studied lower bounds on the error
probability. The random coding bound for MAC was studied by
Gallger~\cite{GalBook}, Pokorny and Wallmeier~\cite{Pokorney}, and
Liu and Hughes~\cite{Liu-RandomCoding}. In this paper, we mostly
concentrate on the result of~\cite{Pokorney}
and~\cite{Liu-RandomCoding}. Both of these random coding theorems
are universal, i.e., a fixed choice of codewords and decoding sets
achieve their upper bounds for all MACs with given input and output
alphabets. In deriving both bounds, three crucial steps are
observed. The first step is the choice of the ensemble.
In~\cite{Pokorney}, each codeword of each code in the ensemble is
chosen from $T_{P_X}$ and $T_{P_Y}$, for some $P_X$ and $P_Y$.
However, in~\cite{Liu-RandomCoding} for a fixed distribution, $P_U
P_{X|U} P_{Y|U}$, the codewords of each code in the ensemble are
chosen from $T_{P_{X|U}}(\textbf{u})$ and $T_{P_{Y|U}}(\textbf{u})$
for some sequence $\textbf{u} \in T_{P_U}$. The second step is the
packing lemma, in which the existence of some particular code with
certain properties is proven. The way the existence of such a code
is proved is through random coding argument over the ensemble. As a
side result of this step, it can be shown that most codes in the
ensemble of~\cite{Pokorney}~\cite{Liu-RandomCoding} have these
properties. In the third step, an appropriate decoding rule is first
chosen, and the performance of the code, found in the packing step,
is analyzed. It has been shown that the result of Liu and Hughes is
tighter than Pokorny's since they used a different ensemble and a
differnet decoding rule. In this work, we follow a similar
three-step approach. First, we start with an ensemble identical
to~\cite{Liu-RandomCoding}. Then, we provide a new packing lemma in
which the resulting code has more constraints in comparison to the
packing lemmas in ~\cite{Pokorney} and~\cite{Liu-RandomCoding}. This
packing lemma is very similar to Pokorny's packing lemma, in the
sense that only channel inputs appear in the packing inequalities.
One of the advantages of this packing lemma, in comparison
to~\cite{Pokorney}, is that it enables us to partially expurgate
some of the codewords and end up with a new code with stronger
properties. In general, expurgation has not been studied in MAC,
since by eliminating some of the codeword pairs, we may end up with
correlated input sequences. In this work, we do not eliminate pairs
of codewords. Rather, we expurgate codewords from only one of the
codebooks. Finally, we analyze the performance of the expurgated
code and end up with a new upper bound on the probability of error.

This paper is organized as follows: section II introduces
terminology, and section III summarizes our main results. The proofs
of some of these results are given in the Appendix.

%
%

\section{Preliminaries}\label{prelim}

For any alphabet $\mathcal{X}$, $\mathcal{P(X)}$ denotes the set of
all probability distributions on $\mathcal{X}$. The \emph{type} of a
sequence $\mathbf{x}=(x_1,...,x_n) \in \mathcal{X}^n$ is the
distributions $P_{\mathbf{x}}$ on $\mathcal{X}$ defined by
\begin{align}
P_{\mathbf{x}}(x)\triangleq
\frac{1}{n}N(x|\mathbf{x}),\;\;\;\;\;\;\;\;\;\;\;\;\;\;\; x \in
\mathcal{X},
\end{align}
where $N(x|\mathbf{x})$ denotes the number of occurrences of $x$ in
$\mathbf{x}$. Let $\mathcal{P}_n (\mathcal{X})$ denote the set of
all types in $\mathcal{X}^n$, and define the set of all sequences in
$\mathcal{X}^n$ of type $P$ as
\begin{align}
T_P \triangleq \{\mathbf{x} \in \mathcal{X}^n: P_{\mathbf{x}}=P\}.
\end{align}
The joint type of a pair $(\mathbf{x},\mathbf{y}) \in \mathcal{X}^n
\times \mathcal{Y}^n$ is the probability distribution
$P_{\mathbf{x},\mathbf{y}}$ on $\mathcal{X} \times \mathcal{Y}$
defined by
\begin{align}
P_{\mathbf{x},\mathbf{y}}(x,y)\triangleq
\frac{1}{n}N(x,y|\mathbf{x},\mathbf{y}),\;\;\;\;\;\;\;\;\;\;\;\;\;\;\;
(x,y) \in \mathcal{X} \times \mathcal{Y},
\end{align}
where $N(x,y|\mathbf{x},\mathbf{y})$ is the number of occurrences of
$(x,y)$ in ($\mathbf{x},\mathbf{y}$). The relative entropy or
\emph{Kullback-Leibler} distance between two probability
distribution $P,Q \; \in \mathcal{P(X)}$ is defined as
\begin{align}
D(P||Q) \triangleq \sum_{x \in
\mathcal{X}}P(x)\log\frac{P(x)}{Q(x)}.
\end{align}
Let $\mathcal{W(Y|X)}$ denote the set of all stochastic matrices
with input alphabet $\mathcal{X}$ and output alphabet $\mathcal{Y}$.
Then, given stochastic matrices $V,\ W \in \mathcal{W(Y|X)}$, the
conditional \emph{I-divergence} is defined by
\begin{align}
D(V||W|P) \triangleq \sum_{x \in
\mathcal{X}}P(x)D(V(\cdot|x)||W(\cdot|x)).
\end{align}
\begin{definition}
An $(n,M,N)$ multi-user code for a given MAC $W$, is a set
$\{(\mathbf{x}_i,\mathbf{y}_j,D_{ij}): 1 \leq i \leq M, 1 \leq j
\leq N\}$ with
\begin{itemize}
\item $\mathbf{x}_i \in \mathcal{X}^n$, $\mathbf{y}_j \in
\mathcal{Y}^n$, $D_{ij} \subset \mathcal{Z}^n$
\item $D_{ij} \cap D_{i'j'}=\varnothing$ for  $(i,j) \neq
(i',j')$.
\end{itemize}
\end{definition}

\begin{definition}
When message $(i,j)$ is transmitted, the conditional probability of
error of the multiuser code $\C$ is given by
\begin{equation*}
e_{ij}(\C,W) \triangleq W^n(D^c_{ij}|\mathbf{x}_i,\mathbf{y}_j).
\end{equation*}
The average probability of error for multiuser code, $\C$, is
defined as
\begin{align}
e(\C,W) \triangleq \frac{1}{M
N}\sum_{i=1}^{M}\sum_{j=1}^{N}e_{ij}(\C,W) .
\end{align}
\end{definition}

\section{main result}
In this section, we present a new, universally achievable upper
bound on the average error probability of multiple-access channel.
We observe that the mutual position of the codewords plays a crucial
role in determining the decoding error. Intuitively, we expect that
the codewords in a ``good'' code must be far from each other. In
accordance with the ideas of Csiszar and Korner~\cite{Csiszarbook},
we use conditional types to quantify this statement. Basically, we
shall select a prescribed number of sequences in $\X^n$ and $\Y^n$
so that the shells around each pair have small intersections with
the shells around other other sequences. In general, we have two
types of packing lemmas based on whether the output of the shell
belongs to the channel input space or channel output space. The
Packing lemma in~\cite{Pokorney} belongs to the first type, and the
one in~\cite{Liu-RandomCoding} belongs to the second type. All the
inequalities in the first type depend only on the channel input
sequences. However, in the second type, the lemma incorporates the
channel output into the packing inequalities. In this work, we use
the first type. In the following, we prove three packing lemmas. In
lemma~\ref{packing1}, we show that there exists a good code with
some certain properties. The nature of these properties is average,
in the sense that they guarantee ,on the average, the codewords in
the code are far from each other.
One can easily show that by using this packing lemma and an
appropriate decoder, all the results of~\cite{Pokorney}
and~\cite{Liu-RandomCoding} can be re-derived and unified. In
lemma~\ref{packing3}, we go one step further, by proving that the
code found in lemma~\ref{packing1} has some additional properties
that are now guaranteed for all individual pairs of sequences. If we
use this packing lemma in bounding the average probability of error,
we will get a tighter bound, especially at low rates. One can show
that most of the random codes from the ensemble have these
properties. Hence, this kind of bound is called the typical random
coding bound in accordance to~\cite{Barg-Random Code}. Finally, In
lemma~\ref{packing2}, we use one of these typical codes and
eliminate some of its codewords. The resulting code has all the
previous properties mentioned in lemma~\ref{packing1} and
lemma~\ref{packing3}. In addition, this code satisfies some
additional stronger constraints.
In lemma~\ref{condtion-lemma}, we show that only some of the joint
type can be seen in the expurgated code. Finally, we calculate  a
new upper bound for the average probability of error, depending only
on the properties of the set of codewords resulting from
expurgation.

\begin{lemma}\label{packing1}
For every finite set $\U$, $P_{XYU} \in \P_n(\U \times \X \times
\Y)$ such that $X-U-Y$, $R_X \geq 0$, $R_Y \geq 0$ , $\delta > 0$,
and $\textbf{u} \in T^n_{P_U}$, there exists sets of codewords
$\C_X=\{\textbf{x}_1,\textbf{x}_2,...,\textbf{x}_{M_X}\}$ and
$\C_Y=\{\textbf{y}_1,\textbf{y}_2,...,\textbf{y}_{M_Y}\}$ with
$\textbf{x}_i \in T^n_{P_{X|U}}(\textbf{u})$, $\textbf{y}_j \in
T^n_{P_{Y|U}}(\textbf{u})$ for all i and j, $M_X \geq 2^{nR_X}$, and
$M_Y \geq 2^{nR_Y}$, such that for every  joint type
$V_{UXY\tilde{X}\tilde{Y}} \in \P_n(\U \times (\X \times \Y)^2)$,
whenever $n \geq n_0(|\U|,|\X|,|\Y|,\delta)$,
\begin{eqnarray}
\frac{1}{M_X M_Y} \sum_{i=1}^{M_X} \sum_{j=1}^{M_Y}
1_{T_{V_{UXY}}}(\textbf{u}, \textbf{x}_i, \textbf{y}_j) \leq
2^{-n[F(V)- 2 \delta]}\label{randomorg1}\;\;\;\;\;\;\;\;\;\;\;\\
\frac{1}{M_X M_Y} \sum_{i=1}^{M_X} \sum_{j=1}^{M_Y} \sum_{l \neq j}
1_{T_{V_{UXY\tilde{Y}}}}(\textbf{u}, \textbf{x}_i, \textbf{y}_j,
\textbf{y}_l)
\nonumber\;\;\;\;\;\;\;\;\;\;\;\;\;\;\;\;\;\;\;\;\;\;\;\\ \leq
2^{-n[F_{Y}(V)- 3 \delta]}\label{randomorg3}\\
\frac{1}{M_X M_Y} \sum_{i=1}^{M_X} \sum_{j=1}^{M_Y} \sum_{k \neq i}
1_{T_{V_{UXY\tilde{X}}}}(\textbf{u}, \textbf{x}_i, \textbf{y}_j,
\textbf{x}_k)
\nonumber\;\;\;\;\;\;\;\;\;\;\;\;\;\;\;\;\;\;\;\;\;\;\\\ \leq
2^{-n[F_{X}(V)- 3
\delta]}\label{randomorg4}\\
\frac{1}{M_X M_Y} \sum_{i=1}^{M_X} \sum_{j=1}^{M_Y} \sum_{k \neq i}
\sum_{l \neq j} 1_{T_{V_{UXY\tilde{X}\tilde{Y}}}}(\textbf{u},
\textbf{x}_i, \textbf{y}_j, \textbf{x}_k, \textbf{y}_l) \nonumber\;\;\;\;\;\;\;\;\;\\
\leq 2^{-n[F_{XY}(V)- 4 \delta]}\label{randomorg2}
\end{eqnarray}
where
\begin{align}
F(V) &\triangleq I_V(X \wedge Y|U)\label{def1}\\
F_X(V) &\triangleq I_V(X \wedge Y|U) + I_V(\tilde{X} \wedge Y|U)
\nonumber\\&+I_V(\tilde{X} \wedge X|UY)-R_X\\
F_Y(V) &\triangleq I_V(X \wedge Y|U) + I_V(X \wedge \tilde{Y}|U)
\nonumber\\&+I_V(\tilde{Y} \wedge Y|UX)-R_Y\\
F_{XY}(V) &\triangleq I_V(X \wedge Y|U) + I_V(\tilde{X} \wedge
\tilde{Y}|U) \nonumber\\&+I_V(\tilde{X}\tilde{Y} \wedge XY|U)-R_X
-R_Y\label{def4}
\end{align}
Here $U$, $X$, $Y$, $\tilde{X}$, $\tilde{Y}$ denote random variables
with common distribution $V_{UXY\tilde{X}\tilde{Y}} \in \P_n(\U
\times (\X \times \Y)^2)$, and $V_{UXY}$, $V_{UXY\tilde{X}}$, and
$V_{UXY\tilde{Y}}$ are appropriate marginal distributions of
$V_{UXY\tilde{X}\tilde{Y}}$.
\end{lemma}
%
%
\begin{proof}
In this proof, we use a similar random coding argument that J.
Pokorny used in \cite{Pokorney}. The main difference is that our
lemma uses a different code ensemble which results in a tighter
bound. Instead of choosing our sequences from $T_{P_X}$ and
$T_{P_Y}$,  we choose our random sequences uniformly from
$T^n_{P_{X|U}}(\textbf{u})$, and $T^n_{P_{Y|U}}(\textbf{u})$ for a
given $\textbf{u} \in T_{P_U}$. In \cite{Liu-RandomCoding}, we see a
similar random code ensemble, however, their packing lemma
incorporates the channel output $\textbf{z}$ into the packing
inequalities. One can easily show that, by using this packing lemma
and considering the minimum equivocation decoding rule, we would end
up with the random coding bound derived in~\cite{Liu-RandomCoding}.
\end{proof}


\begin{lemma}\label{packing3}
For every finite set $\U$, $P_{XYU} \in \P_n(\U \times \X \times
\Y)$ such that $X-U-Y$, $R_X \geq 0$, $R_Y \geq 0$ , $\delta > 0$,
and $\textbf{u} \in T^n_{P_U}$, there exists sets of codewords
$\C_X=\{\textbf{x}_1,\textbf{x}_2,...,\textbf{x}_{M_X}\}$ and
$\C_Y=\{\textbf{y}_1,\textbf{y}_2,...,\textbf{y}_{M_Y}\}$ with
$\textbf{x}_i \in T^n_{P_{X|U}}(\textbf{u})$, $\textbf{y}_j \in
T^n_{P_{Y|U}}(\textbf{u})$ for all i and j, $M_X \geq 2^{nR_X}$, and
$M_Y \geq 2^{nR_Y}$, such that for every  joint type
$V_{UXY\tilde{X}\tilde{Y}} \in \P_n(\U \times (\X \times \Y)^2)$,
~\eqref{randomorg1}-~\eqref{randomorg2} are satisfied provided $n
\geq n_0(|\U|,|\X|,|\Y|,\delta)$. Moreover, for any $1 \leq i \leq
M_X$, and any $1 \leq j \leq M_Y$
\begin{eqnarray}
 1_{T_{V_{UXY}}}(\textbf{u},
\textbf{x}_i, \textbf{y}_j)
\leq 2^{-n[F(V) - R_X - R_Y - 2 \delta]} \;\; \label{randomtypical}\;\;\;\;\;\;\;\;\;\;\;\;\;\;\;\;\;\;\\
\sum_{k \neq i} 1_{T_{V_{UXY\tilde{X}}}}(\textbf{u}, \textbf{x}_i,
\textbf{y}_j, \textbf{x}_k) \leq 2^{-n[F_{X}(V)- R_X -R_Y - 3
\delta]}\;\;\;\;\;\;\;\label{randomtypical2}\\
\sum_{l \neq j}  1_{T_{V_{UXY\tilde{Y}}}}(\textbf{u}, \textbf{x}_i,
\textbf{y}_j, \textbf{y}_l) \leq 2^{-n[F_{Y}(V)- R_X - R_Y -3
\delta]}\;\;\;\;\;\;\;\label{randomtypical3}\\
\sum_{k \neq i} \sum_{l \neq j}
1_{T_{V_{UXY\tilde{X}\tilde{Y}}}}(\textbf{u}, \textbf{x}_i,
\textbf{y}_j, \textbf{x}_k, \textbf{y}_l)
\;\;\;\;\;\;\;\;\;\;\;\;\;\;\;\;\;\;\;\;\;\;\;\;\;\;\;\;\;\;\;\;\;\;\nonumber
\\\leq 2^{-n[F_{XY}(V)-  R_X - R_Y- 4
\delta]}\label{randomtypical4},
\end{eqnarray}
\end{lemma}
\begin{proof}
Let us use the result of lemma~\ref{packing1}, and multiply both
sides of the inequalities~\eqref{randomorg1}-~\eqref{randomorg2} by
$M_X M_Y$.
\end{proof}
%
%
\begin{lemma}\label{packing2}
For every finite set $\U$, $P_{XYU} \in \P_n(\U \times \X \times
\Y)$ such that $X-U-Y$, $R_X \geq 0$, $R_Y \geq 0$ , $\delta > 0$,
and $\textbf{u} \in T^n_{P_U}$, there exists sets of codewords
$\C^*_X=\{\textbf{x}_1,\textbf{x}_2,...,\textbf{x}_{M^*_X}\}$ and
$\C^*_Y=\{\textbf{y}_1,\textbf{y}_2,...,\textbf{y}_{M^*_Y}\}$ with
$\textbf{x}_i \in T^n_{P_{X|U}}$, $\textbf{y}_j \in T^n_{P_{Y|U}}$
for all i and j, $M^*_X \geq 2^{n(R_X-\delta)}$, and $M^*_Y \geq
2^{n(R_Y- \delta)}$, such that for every  joint type
$V_{UXY\tilde{X}\tilde{Y}} \in \P_n(\U \times (\X \times \Y)^2)$,
\begin{eqnarray}
\frac{1}{M^*_X M^*_Y} \sum_{i=1}^{M^*_X} \sum_{j=1}^{M^*_Y}
1_{T_{V_{UXY}}}(\textbf{u}, \textbf{x}_i, \textbf{y}_j) \leq
2^{-n[F(V)- 3 \delta]}\;\\
\frac{1}{M^*_X M^*_Y} \sum_{i=1}^{M^*_X} \sum_{j=1}^{M_Y} \sum_{k
\neq i} 1_{T_{V_{UXY\tilde{X}}}}(\textbf{u}, \textbf{x}_i,
\textbf{y}_j, \textbf{x}_k) \nonumber\;\;\;\;\;\;\;\;\;\;\;\;\\ \leq
2^{-n[F_{X}(V)- 4
\delta]}\\
\frac{1}{M^*_X M^*_Y} \sum_{i=1}^{M^*_X} \sum_{j=1}^{M^*_Y} \sum_{l
\neq j} 1_{T_{V_{UXY\tilde{Y}}}}(\textbf{u}, \textbf{x}_i,
\textbf{y}_j, \textbf{y}_l) \nonumber\;\;\;\;\;\;\;\;\;\;\;\;\;\\
\leq 2^{-n[F_{Y}(V)- 4
\delta]}\\
\frac{1}{M^*_X M^*_Y} \sum_{i=1}^{M^*_X} \sum_{j=1}^{M^*_Y} \sum_{k
\neq i} \sum_{l \neq j}
1_{T_{V_{UXY\tilde{X}\tilde{Y}}}}(\textbf{u},
\textbf{x}_i, \textbf{y}_j, \textbf{x}_k, \textbf{y}_l) \nonumber\\
\leq 2^{-n[F_{XY}(V)- 5 \delta]}
\end{eqnarray}
and for any $1 \leq i \leq M^*_X$, and any $1 \leq j \leq M^*_Y$
\begin{eqnarray}
 1_{T_{V_{UXY}}}(\textbf{u},
\textbf{x}_i, \textbf{y}_j)
\leq 2^{-n[F(V) -\min\{R_X,R_Y\} - 3 \delta]} \;\; \label{randomnew1}\;\;\;\;\;\;\;\;\;\;\;\;\\
\sum_{k \neq i} 1_{T_{V_{UXY\tilde{X}}}}(\textbf{u}, \textbf{x}_i,
\textbf{y}_j, \textbf{x}_k) \leq 2^{-n[F_{X}(V)- \min\{R_X,R_Y\} - 4
\delta]}\label{randomnew2}\\
\sum_{l \neq j}  1_{T_{V_{UXY\tilde{Y}}}}(\textbf{u}, \textbf{x}_i,
\textbf{y}_j, \textbf{y}_l) \leq 2^{-n[F_{Y}(V)- \min\{R_X,R_Y\}-4
\delta]}\label{randomnew3}\\
\sum_{k \neq i} \sum_{l \neq j}
1_{T_{V_{UXY\tilde{X}\tilde{Y}}}}(\textbf{u}, \textbf{x}_i,
\textbf{y}_j, \textbf{x}_k, \textbf{y}_l)
\;\;\;\;\;\;\;\;\;\;\;\;\;\;\;\;\;\;\;\;\;\;\;\;\;\;\;\;\;\;\;\;\;\;\nonumber
\\\leq 2^{-n[F_{XY}(V)- \min\{R_X,R_Y\} \min\{R_X,R_Y\}- 5
\delta]}\label{randomnew4},
\end{eqnarray}
whenever
\begin{equation*}
n \geq n_0(|\U|,|\X|,|\Y|,\delta)
\end{equation*}
where $F(V), F_X(V), F_Y(V), F_{XY}(V)$ are defined in
\eqref{def1}-\eqref{def4}.
\end{lemma}
\begin{proof}
Let $\C_X=\{\textbf{x}_1,\textbf{x}_2,...,\textbf{x}_{M_X}\}$ and
$\C_Y=\{\textbf{y}_1,\textbf{y}_2,...,\textbf{y}_{M_Y}\}$ be the
collections of codewords whose existence is asserted in
lemma~\ref{packing1}. From lemma~\ref{packing1}, the codewords
satisfy
\begin{eqnarray}
\frac{1}{M_Y} \sum_{j=1}^{M_Y} \frac{1}{M_X}\sum_{i=1}^{M_X}
1_{T_{V_{UXY}}}(\textbf{u}, \textbf{x}_i, \textbf{y}_j) \leq
2^{-n[F(V)- 2 \delta]}.
\end{eqnarray}
Therefore, there exist $M_Y^1 \geq \frac{M_Y}{2}$ codewords in
$\C_Y$ that satisfy
\begin{eqnarray}
\frac{1}{M_X}\sum_{i=1}^{M_X} 1_{T_{V_{UXY}}}(\textbf{u},
\textbf{x}_i, \textbf{y}_j) \leq 2^{-n[F(V)- 2 \delta]} \times 2
\label{lemma-2-prf-form2}.
\end{eqnarray}
Let us  call this set of codewords $\C^1_Y$. By multiplying both
sides of~\eqref{lemma-2-prf-form2} with $M_X$, and considering the
fact that all terms in the summation are nonnegative, it can be
concluded that for every $\textbf{x}_i \in \C_X$, $\textbf{y}_j \in
\C^1_Y$,
\begin{eqnarray}
1_{T_{V_{UXY}}}(\textbf{u}, \textbf{x}_i, \textbf{y}_j)  \leq
2^{-n[F(V)- 2 \delta -R_X]} \times 2 \label{randomeq1}.
\end{eqnarray}
 We can make a similar argument and conclude that there
exists a subset of $\C_X$, called $\C^1_X$, with $M_X^1 \geq
\frac{M_X}{2}$ codewords such that for any $\textbf{x}_i \in
\C^1_X$, $\textbf{y}_j \in \C_Y$
\begin{eqnarray}
1_{T_{V_{UXY}}}(\textbf{u}, \textbf{x}_i, \textbf{y}_j)  \leq
2^{-n[F(V)- 2 \delta -R_Y]} \times 2.
\end{eqnarray}
Without loss of generality, let us assume $R_X < R_Y$. In this case,
\eqref{randomeq1} will end up with a tighter result.
Using~\eqref{randomorg3}, we conclude that
\begin{equation*}
\frac{1}{M_X M_Y} \sum_{\substack{i \in \C_X\\ j \in \C^1_Y}}
\sum_{l \neq j} 1_{T_{V_{UXY\tilde{Y}}}}(\textbf{u}, \textbf{x}_i,
\textbf{y}_j, \textbf{y}_l) \nonumber\\ \leq 2^{-n[F_{Y}(V)- 3
\delta]}.
\end{equation*}
Since $M_Y^1 \geq \frac{M_Y}{2}$,
\begin{equation*}
\frac{1}{M_X M^1_Y} \sum_{\substack{i \in \C_X\\ j \in \C^1_Y}}
\sum_{l \neq j} 1_{T_{V_{UXY\tilde{Y}}}}(\textbf{u}, \textbf{x}_i,
\textbf{y}_j, \textbf{y}_l) \nonumber\\ \leq 2^{-n[F_{Y}(V)- 3
\delta]} \times 2,
\end{equation*}
again, by a similar argument, there exists $M_Y^2 \geq
\frac{M^1_Y}{2}$ codewords, $\textbf{y}_j$, in $\C^1_Y$ such that
\begin{eqnarray}
\frac{1}{M_X}\sum_{i=1}^{M_X} \sum_{l \neq j}
1_{T_{V_{UXY\tilde{Y}}}}(\textbf{u}, \textbf{x}_i, \textbf{y}_j,
\textbf{y}_l) \nonumber\\ \leq 2^{-n[F_{Y}(V)- 3 \delta]} \times 4.
\end{eqnarray}
Let us call this subset of $\C^1_Y$ as $\C^2_Y$. Therefore, for any
$\textbf{x}_i \in \C_X$, $\textbf{y}_j \in \C^2_Y$,
\begin{eqnarray}
\sum_{l \neq j}  1_{T_{V_{UXY\tilde{Y}}}}(\textbf{u}, \textbf{x}_i,
\textbf{y}_j, \textbf{y}_l) \leq 2^{-n[F_{Y}(V)- 3 \delta-R_X]}
\times 4. \label{randomeq2}
\end{eqnarray}
By using \eqref{randomorg4}, we can conclude that
\begin{equation*}
\frac{1}{M_X M_Y} \sum_{\substack{i \in \C_X\\ j \in \C^2_Y}}
\sum_{k \neq i} 1_{T_{V_{UXY\tilde{X}}}}(\textbf{u}, \textbf{x}_i,
\textbf{y}_j, \textbf{x}_k) \nonumber\\ \leq 2^{-n[F_{X}(V)- 3
\delta]}
\end{equation*}
Considering the fact that $M^2_Y \geq \frac{M^1_Y}{2} \geq
\frac{M_Y}{4}$, we can conclude that
\begin{equation}
\frac{1}{M_X M^2_Y} \sum_{\substack{i \in \C_X\\ j \in \C^2_Y}}
\sum_{k \neq i} 1_{T_{V_{UXY\tilde{X}}}}(\textbf{u}, \textbf{x}_i,
\textbf{y}_j, \textbf{x}_k) \nonumber\\ \leq 2^{-n[F_{X}(V)- 3
\delta]} \times 4
\end{equation}
Hence, there exists  $C^3_Y \subset C^2_Y$, with $M^3_Y \geq
\frac{M^2_Y}{2}$ codewords such that
\begin{eqnarray}
\frac{1}{M_X}  \sum_{\substack{i \in \C_X}} \sum_{k \neq i}
1_{T_{V_{UXY\tilde{X}}}}(\textbf{u}, \textbf{x}_i, \textbf{y}_j,
\textbf{x}_k) \nonumber\\ \leq 2^{-n[F_{X}(V)- 3 \delta]} \times 8
\end{eqnarray}
Therefore, for all $\textbf{x}_i \in \C_X$, $\textbf{y}_j \in
\C^3_Y$,
\begin{eqnarray}
\sum_{k \neq i} 1_{T_{V_{UXY\tilde{X}}}}(\textbf{u}, \textbf{x}_i,
\textbf{y}_j, \textbf{x}_k) \leq 2^{-n[F_{X}(V)- R_X - 3 \delta]}
\times 8 \label{randomeq3}.
\end{eqnarray}
Similarly, by using \eqref{randomorg2}, we can conclude that
\begin{eqnarray}
\frac{1}{M_X M_Y} \sum_{\substack{i \in \C_X\\ j \in \C^3_Y}}
\sum_{k \neq i} \sum_{l \neq j}
1_{T_{V_{UXY\tilde{X}\tilde{Y}}}}(\textbf{u}, \textbf{x}_i,
\textbf{y}_j, \textbf{x}_k, \textbf{y}_l) \nonumber\\ \leq
2^{-n[F_{XY}(V)- 3 \delta]}.
\end{eqnarray}
By a similar argument and using the fact that $M^3_Y \geq
\frac{M^2_Y}{2} \geq \frac{M^1_Y}{4} \geq \frac{M_Y}{8}$, we
conclude that
\begin{eqnarray}
\frac{1}{M_X M^3_Y} \sum_{\substack{i \in \C_X\\ j \in \C^3_Y}}
\sum_{k \neq i} \sum_{l \neq j}
1_{T_{V_{UXY\tilde{X}\tilde{Y}}}}(\textbf{u}, \textbf{x}_i,
\textbf{y}_j, \textbf{x}_k, \textbf{y}_l) \nonumber\\ \leq
2^{-n[F_{XY}(V)- 3 \delta]} \times 8.
\end{eqnarray}
Therefore, there exist $\C^4_Y \subset \C^3_Y$, with $M^4_Y \geq
\frac{M^3_Y}{2}$ codewords, such that
\begin{eqnarray}
\frac{1}{M_X} \sum_{\substack{i \in \C_X}} \sum_{k \neq i} \sum_{l
\neq j} 1_{T_{V_{UXY\tilde{X}\tilde{Y}}}}(\textbf{u}, \textbf{x}_i,
\textbf{y}_j, \textbf{x}_k, \textbf{y}_l) \nonumber\\ \leq
2^{-n[F_{XY}(V)- 3 \delta]} \times 16.
\end{eqnarray}
Similarly, since  $M^4_Y \geq \frac{M^3_Y}{2} \geq \frac{M^2_Y}{4}
\geq \frac{M^1_Y}{8} \geq \frac{M_Y}{16}$, we conclude that for all
$\textbf{x}_i \in \C_X$, $\textbf{y}_j \in \C^4_Y$,
\begin{eqnarray}
\sum_{k \neq i} \sum_{l \neq j}
1_{T_{V_{UXY\tilde{X}\tilde{Y}}}}(\textbf{u}, \textbf{x}_i,
\textbf{y}_j, \textbf{x}_k, \textbf{y}_l)
\;\;\;\;\;\;\;\;\;\;\;\;\;\;\;\;\;\;\;\;\;\nonumber
\\\leq 2^{-n[F_{XY}(V)- R_X - 3 \delta]} \times 16
\label{randomeq4}.
\end{eqnarray}
Since  $\C^4_Y \subset \C^3_Y \subset \C^2_Y \subset \C^1_Y$, any
codeword belonging to $C^4_Y$ has all the properties we derived in
\eqref{randomeq1}, \eqref{randomeq2}, \eqref{randomeq3},
\eqref{randomeq4}. Therefore, we have proven that there exists a
codebook $\C^4_Y \subset C_Y$ with $M^4_Y \geq \frac{M_Y}{16}$
codewords such that for any $\textbf{x}_i \in \C_X$, $\textbf{y}_j
\in \C^4_Y$, we have the properties \eqref{randomeq1},
\eqref{randomeq2}, \eqref{randomeq3}, \eqref{randomeq4}. As shown,
we have eliminated some of the codewords from $\C_Y$. Similarly, we
can do the expurgation on $\C_X$. If $R_X < R_Y$, the expurgation on
$\C_Y$ results in a tighter result. However, if $R_X > R_Y$, the
expurgation on $\C_X$ would end up with a tighter bound. Thus, in
general, there exists a pair of codebooks ($\C^*_X$, $\C^*_Y$), with
$|\C^*_X||\C^*_Y| \geq \frac{|\C_X||\C_Y|}{16}$, such that for any
$\textbf{x}_i \in \C^*_X$, $\textbf{y}_j \in \C^*_Y$,
\begin{eqnarray}
1_{T_{V_{UXY}}}(\textbf{u}, \textbf{x}_i, \textbf{y}_j)  \leq
2^{-n[F(V) -\min\{R_X,R_Y\} - 3 \delta]} \label{type1}\;\;\;\;\;\;\;\;\;\;\;\;\;\;\; \\
\sum_{k \neq i} 1_{T_{V_{UXY\tilde{X}}}}(\textbf{u}, \textbf{x}_i,
\textbf{y}_j, \textbf{x}_k) \leq 2^{-n[F_{X}(V)- \min\{R_X,R_Y\} - 4
\delta]}\label{type2}\\
\sum_{l \neq j}  1_{T_{V_{UXY\tilde{Y}}}}(\textbf{u}, \textbf{x}_i,
\textbf{y}_j, \textbf{y}_l) \leq 2^{-n[F_{Y}(V)- \min\{R_X,R_Y\}-4
\delta]}\label{type3}\\
\sum_{k \neq i} \sum_{l \neq j}
1_{T_{V_{UXY\tilde{X}\tilde{Y}}}}(\textbf{u}, \textbf{x}_i,
\textbf{y}_j, \textbf{x}_k, \textbf{y}_l)
\;\;\;\;\;\;\;\;\;\;\;\;\;\;\;\;\;\;\;\;\;\;\;\;\;\;\;\;\;\;\;\;\;\;\nonumber
\\\leq 2^{-n[F_{XY}(V)- \min\{R_X,R_Y\} \min\{R_X,R_Y\}- 5
\delta]}\label{type4}
\end{eqnarray}
The only difference between the exponents in
\eqref{randomnew1}-\eqref{randomnew4} and the ones in
\eqref{randomorg1}-\eqref{randomorg2} is $\min\{R_X,R_Y\}$. Despite
of the \eqref{randomorg1}-\eqref{randomorg2} which are upper bounds
for some quantities averaged over all pairs of sequences belonging
to $(\C_X,\C_Y)$, the results in
\eqref{randomnew1}-\eqref{randomnew4} are valid for all pairs of
codewords in ($\C^*_X$,$\C^*_Y$). Let us define $M^*_X \triangleq
|\C^*_X|$, $M^*_Y \triangleq |\C^*_Y|$. In the following, we will
show that the new codebook pair, ($\C^*_X$,$\C^*_Y$), still
satisfies the same average performance bound we obtained for the
original codebook pair, $(\C_X,\C_Y)$. The functions in
\eqref{randomorg1}-\eqref{randomorg2} for the new codebook pair can
be upperbounded as follows,
\begin{eqnarray}
& &\frac{1}{M^*_X M^*_Y} \sum_{i=1}^{M^*_X} \sum_{j=1}^{M^*_Y}
1_{T_{V_{UXY}}}(\textbf{u}, \textbf{x}_i, \textbf{y}_j) \nonumber \\
\nonumber&\leq& \frac{1}{M^*_X M^*_Y} \sum_{i=1}^{M_X}
\sum_{j=1}^{M_Y} 1_{T_{V_{UXY}}}(\textbf{u}, \textbf{x}_i,
\textbf{y}_j) \\ \nonumber &\leq& \frac{16}{M_X M_Y}
\sum_{i=1}^{M_X} \sum_{j=1}^{M_Y} 1_{T_{V_{UXY}}}(\textbf{u},
\textbf{x}_i, \textbf{y}_j) \nonumber \\ &\leq& 16
* 2^{-n[F(V)- 3 \delta]}  \leq 2^{-n[F(V)- 2 \delta]}.
\end{eqnarray}
We can use a similar argument and show that
\begin{align}
&\frac{1}{M^*_X M^*_Y} \sum_{i=1}^{M^*_X} \sum_{j=1}^{M^*_Y} \sum_{k
\neq i} \sum_{l \neq j}
1_{T_{V_{UXY\tilde{X}\tilde{Y}}}}(\textbf{u},
\textbf{x}_i, \textbf{y}_j, \textbf{x}_k, \textbf{y}_l) \nonumber\\
&\;\;\;\;\;\;\;\;\;\;\;\;\;\;\;\;\;\;\;\;\;\;\;\;\;\;\;\;\;\;\;\;\;\;\;\;\;\;\;\;\;\;\;\;\;\leq
2^{-n[F_{XY}(V)- 4 \delta]}\\
&\frac{1}{M^*_X M^*_Y} \sum_{i=1}^{M^*_X} \sum_{j=1}^{M^*_Y} \sum_{l
\neq j} 1_{T_{V_{UXY\tilde{Y}}}}(\textbf{u}, \textbf{x}_i,
\textbf{y}_j, \textbf{y}_l) \nonumber\\
&\;\;\;\;\;\;\;\;\;\;\;\;\;\;\;\;\;\;\;\;\;\;\;\;\;\;\;\;\;\;\;\;\;\;\;\;\;\;\;\;\;\;\;\;\;\leq
2^{-n[F_{Y}(V)- 4 \delta]}\\
&\frac{1}{M^*_X M^*_Y} \sum_{i=1}^{M^*_X} \sum_{j=1}^{M^*_Y} \sum_{k
\neq i} 1_{T_{V_{UXY\tilde{X}}}}(\textbf{u}, \textbf{x}_i,
\textbf{y}_j, \textbf{x}_k) \nonumber\\
&\;\;\;\;\;\;\;\;\;\;\;\;\;\;\;\;\;\;\;\;\;\;\;\;\;\;\;\;\;\;\;\;\;\;\;\;\;\;\;\;\;\;\;\;\;\leq
2^{-n[F_{X}(V)- 5 \delta]}
\end{align}
Here, by method of expurgation, we end up with a code with a similar
average bound as we had for the original code. However, all pairs of
codewords in the new code also satisfy
\eqref{randomnew1}-\eqref{randomnew4}. Therefore, we did not lose
anything in terms of average performance, however, as we see in
theorem~\ref{randomcodingthm} , we would end up with a tighter
random coding bound since we have more constraints on any particular
pair of codewords in our codebook pair.
\end{proof}
\begin{lemma}\label{condtion-lemma}
For any type $V_{UXY\tilde{X}\tilde{Y}} \in \P_n(\U \times (\X
\times \Y)^2)$ such that for some $\textbf{x}_i, \textbf{x}_k \in
C^*_X$, and $\textbf{y}_j, \textbf{y}_l \in C^*_Y$,
\begin{equation}
(\textbf{u}, \textbf{x}_i, \textbf{y}_j, \textbf{x}_k, \textbf{y}_l)
\in T_{V_{UXY\tilde{X}\tilde{Y}}}
\end{equation}
the following inequalities must be satisfied
\begin{align}\label{conditions}
&V_{XU}=V_{\tilde{X}U}=P_{XU}, V_{YU}=V_{\tilde{Y}U}=P_{YU}\nonumber\\
&I_V(X \wedge Y| U), I_V(X \wedge \tilde{Y}| U)\leq \min\{R_X,R_Y\} + 3 \delta\nonumber\\
&I_V(\tilde{X} \wedge Y| U),I_V(\tilde{X} \wedge \tilde{Y}| U)\leq \min\{R_X,R_Y\}  + 3 \delta\nonumber\\
&I_V(X \wedge Y|U) + I_V(\tilde{X} \wedge Y|U) +I_V(\tilde{X} \wedge
X|UY)\nonumber \\ &\;\;\;\;\;\;\;\;\;\;\;\;\;\;\;\;\;\;\;\;\;\;\;\;\;\;\;\leq R_X +\min\{R_X,R_Y\} + 4 \delta \nonumber\\
&I_V(X \wedge \tilde{Y}|U) + I_V(\tilde{X} \wedge \tilde{Y}|U)
+I_V(\tilde{X} \wedge
X|U\tilde{Y})\nonumber \\ &\;\;\;\;\;\;\;\;\;\;\;\;\;\;\;\;\;\;\;\;\;\;\;\;\;\;\;\leq R_X +\min\{R_X,R_Y\} + 4 \delta \nonumber\\
&I_V(X \wedge Y|U) + I_V(X \wedge \tilde{Y}|U) +I_V(\tilde{Y} \wedge
Y|UX)\nonumber \\ &\;\;\;\;\;\;\;\;\;\;\;\;\;\;\;\;\;\;\;\;\;\;\;\;\;\;\;\leq R_Y +\min\{R_X,R_Y\} + 4 \delta\nonumber\\
&I_V(\tilde{X} \wedge Y|U) + I_V(\tilde{X} \wedge \tilde{Y}|U)
+I_V(\tilde{Y} \wedge
Y|U\tilde{X})\nonumber \\ &\;\;\;\;\;\;\;\;\;\;\;\;\;\;\;\;\;\;\;\;\;\;\;\;\;\;\;\leq R_Y +\min\{R_X,R_Y\} + 4 \delta \nonumber\\
&I_V(X \wedge Y|U) + I_V(\tilde{X} \wedge \tilde{Y}|U)
+I_V(\tilde{X}\tilde{Y} \wedge
XY|U)\nonumber \\ &\;\;\;\;\;\;\;\;\;\;\;\;\;\;\;\;\;\;\;\leq R_X+R_Y +\min\{R_X,R_Y\} + 5 \delta \nonumber\\
&I_V(\tilde{X} \wedge Y|U) + I_V(X \wedge \tilde{Y}|U)
+I_V(X\tilde{Y} \wedge \tilde{X}Y|U)\nonumber \\
&\;\;\;\;\;\;\;\;\;\;\;\;\;\;\;\;\;\;\;\leq R_X+R_Y +\min\{R_X,R_Y\}
+ 5 \delta
\end{align}
\end{lemma}
\begin{proof}
Let $(C^*_X,C^*_Y)$ be the collections of codewords whose existence
is asserted in lemma~\ref{packing2}. Consider any $\textbf{x}_i,
\textbf{x}_k \in C^*_X$, and $\textbf{y}_j, \textbf{y}_l \in C^*_Y$.
Let us call their joint empirical distribution of ($\textbf{u},
\textbf{x}_i, \textbf{y}_j, \textbf{x}_k, \textbf{y}_l$) as
$V_{UXY\tilde{X}\tilde{Y}} (u,x,y,\tilde{x},\tilde{y})$. Using
~\eqref{type1}, and the fact that $(\textbf{u}, \textbf{x}_i,
\textbf{y}_j) \in T_{V_{UXY}}$,
\begin{eqnarray}
1 \leq 2^{-n[F(V) -\min\{R_X,R_Y\} - 3 \delta]}
\end{eqnarray}
Therefore,
\begin{eqnarray}
I_V(X \wedge Y|U) \leq \min\{R_X,R_Y\} + 3 \delta \label{prop1-1}
\end{eqnarray}
Similarly, using the empirical distribution of ($\textbf{u},
\textbf{x}_i, \textbf{y}_l$), ($\textbf{u}, \textbf{x}_k,
\textbf{y}_j$), and ($\textbf{u}, \textbf{x}_k, \textbf{y}_l$), we
conclude that
\begin{eqnarray}
I_V(\tilde{X} \wedge Y|U) \leq \min\{R_X,R_Y\} + 3 \delta\label{prop1-2}\\
I_V(X \wedge \tilde{Y}|U) \leq \min\{R_X,R_Y\} + 3 \delta\\
I_V(\tilde{X} \wedge \tilde{Y}|U) \leq \min\{R_X,R_Y\} + 3 \delta
\end{eqnarray}
Since $(\textbf{u}, \textbf{x}_i, \textbf{y}_j, \textbf{x}_k) \in
T_{V_{UXY\tilde{X}}}$,
\begin{eqnarray}
1 \leq \sum_{k \neq i} 1_{T_{V_{UXY\tilde{X}}}}(\textbf{u},
\textbf{x}_i, \textbf{y}_j, \textbf{x}_k) \label{45}
\end{eqnarray}
Using~\eqref{45}, and the upper bound we obtained in ~\eqref{type2},
\begin{eqnarray}
I_V(X \wedge Y|U) + I_V(\tilde{X} \wedge Y|U) +I_V(\tilde{X} \wedge
X|UY)\nonumber \\ \leq R_X +\min\{R_X,R_Y\} + 4 \delta
\label{prop1-3}
\end{eqnarray}
Similarly, since $(\textbf{u}, \textbf{x}_i, \textbf{y}_l,
\textbf{x}_k) \in T_{V_{UX\tilde{Y}\tilde{X}}}$, we conclude that
\begin{eqnarray}
I_V(X \wedge \tilde{Y}|U) + I_V(\tilde{X} \wedge \tilde{Y}|U)
+I_V(\tilde{X} \wedge X|U\tilde{Y})\nonumber \\ \leq R_X
+\min\{R_X,R_Y\} + 4 \delta
\end{eqnarray}
By a similar argument for the empirical distribution of
$(\textbf{u}, \textbf{x}_i, \textbf{y}_j, \textbf{y}_l)$,
$(\textbf{u}, \textbf{x}_k, \textbf{y}_j, \textbf{y}_l)$, and using
the upper bound we obtained in~\eqref{type3}, the following would
respectively be concluded
\begin{eqnarray}
I_V(X \wedge Y|U) + I_V(X \wedge \tilde{Y}|U) +I_V(\tilde{Y} \wedge
Y|UX)\nonumber \\ \leq R_Y +\min\{R_X,R_Y\} + 4 \delta\\
I_V(\tilde{X} \wedge Y|U) + I_V(\tilde{X} \wedge \tilde{Y}|U)
+I_V(\tilde{Y} \wedge Y|U\tilde{X})\nonumber \\ \leq R_Y
+\min\{R_X,R_Y\} + 4 \delta
\end{eqnarray}
Finally, using the empirical distribution of $(\textbf{u},
\textbf{x}_i, \textbf{y}_j, \textbf{x}_k, \textbf{y}_l)$,
$(\textbf{u}, \textbf{x}_k, \textbf{y}_j, \textbf{x}_i,
\textbf{y}_l)$, and the upper bound in~\eqref{type4},
\begin{eqnarray}
I_V(X \wedge Y|U) + I_V(\tilde{X} \wedge \tilde{Y}|U)
+I_V(\tilde{X}\tilde{Y} \wedge XY|U) \nonumber \\\leq R_X + R_Y +
+\min\{R_X,R_Y\} + 5 \delta\\
I_V(\tilde{X} \wedge Y|U) + I_V(X \wedge \tilde{Y}|U)
+I_V(X\tilde{Y} \wedge \tilde{X}Y|U) \nonumber \\\leq R_X + R_Y +
+\min\{R_X,R_Y\} + 5 \delta.
\end{eqnarray}
\end{proof}
\begin{theorem}\label{randomcodingthm}
For every finite set $\U$, $\P_{XYU} \in \P_n(\X \times \Y \times
\U)$ such that $X-U-Y$ , $R_X \geq 0$, $R_Y \geq 0$ , $\delta > 0$,
and $\textbf{u} \in T^n_{P_U}$, there exists a multi-user code
\begin{equation}
\C = \{ (\textbf{x}_i, \textbf{y}_j,D_{ij}) : i=1,...M^*_X,
j=1,...M^*_Y \}
\end{equation}
with $\textbf{x}_i \in T_{P_{X|U}}(\textbf{u})$, $\textbf{y}_j \in
T_{P_{Y|U}}(\textbf{u})$ for all $i$ and $j$, $M^*_X \geq
2^{n(R_X-\delta)}$, and $M^*_Y \geq 2^{n(R_Y-\delta)}$, such that
for every MAC $W:\X \times \Y \rightarrow \Z$
\begin{equation}
e(\C,W) \leq 2^{-n[E_{ex}(R_X,R_Y,W,P_{XYU})-
\delta]}\label{expur-bnd}
\end{equation}
whenever $n \geq n_1(|\Z|,|\X|,|\Y|,|\U|, \delta)$, where
\begin{eqnarray}
E_{ex}(R_X,R_Y,W,P_{XYU}) && \nonumber\\
\triangleq \min_{\substack{\beta=X,Y,XY}}
&&E_{\beta}(R_X,R_Y,W,P_{XYU})\label{Expuragted-def}
\end{eqnarray}
and $E_{\beta}(R_X,R_Y,W,P_{XYU})$, $\beta=X,Y,XY$ are defined
respectively by
\begin{align}
&E_{X}(R_X,R_Y,W,P_{XYU}) \triangleq \nonumber\\
 &\min_{\substack{V_{UXY\tilde{X}Z}} \in \V_X} D(V_{Z|XYU} || W |P_{XYU}) + I_V(X \wedge Y |U) \nonumber\\&+ |I(\tilde{X} \wedge XZ |YU) + I_V(\tilde{X} \wedge Y| U)- R_X|^+ \label{Ex-def}\\ \nonumber\\
&E_{Y}(R_X,R_Y,W,P_{XYU}) \triangleq \nonumber\\
 &\min_{\substack{V_{UXY\tilde{Y}Z}} \in \V_Y} D(V_{Z|XYU} || W |P_{XYU}) + I_V(X \wedge Y |U) \nonumber\\&+ |I(\tilde{Y} \wedge YZ |XU) + I_V(X \wedge \tilde{Y}| U)- R_Y|^+ \label{Ey-def}\\ \nonumber\\
&E_{XY}(R_X,R_Y,W,P_{XYU}) \triangleq \nonumber\\
 &\min_{\substack{V_{UXY\tilde{X}\tilde{Y}Z}} \in \V_{XY}} D(V_{Z|XYU} || W |P_{XYU}) + I_V(X \wedge Y |U) \nonumber\\&+ |I(\tilde{X}\tilde{Y} \wedge XYZ |U) + I_V(\tilde{X} \wedge \tilde{Y}| U)- R_X - R_Y|^+ \label{Exy-def}
\end{align}
where
\begin{align}
&\V_{X} \triangleq \{V_{UXY\tilde{X}Z} :\alpha(V_{UXYZ}) \geq \alpha(V_{U\tilde{X}YZ})\nonumber\\
&\text{and } V_{UXY\tilde{X}} \text{satisfies the relevant
conditions in Lemma \ref{condtion-lemma}} \}\nonumber\\
&\V_{Y} \triangleq \{V_{UXY\tilde{Y}Z} :\alpha(V_{UXYZ}) \geq \alpha(V_{UX\tilde{Y}Z})\nonumber\\
&\text{and } V_{UXY\tilde{Y}} \text{satisfies the relevant
conditions in Lemma \ref{condtion-lemma}} \}\nonumber\\
&\V_{XY} \triangleq \{V_{UXY\tilde{X}\tilde{Y}Z} :\alpha(V_{UXYZ}) \geq \alpha(V_{U\tilde{X}\tilde{Y}Z})\nonumber\\
&\text{and } V_{UXY\tilde{X}\tilde{Y}} \text{satisfies all the
conditions in Lemma \ref{condtion-lemma}} \}
\end{align}
\end{theorem}
\begin{remark}
This exponential error bound can be universally obtained for all
MAC's with given input and output alphabets. Note, it is a universal
bound since the choice of the codewords does not depend on the
channel, and the decoding rule is independent of the channel
statistics.
\end{remark}
\begin{proof}
Fix $\U$, $\P_{XYU} \in \P_n(\X \times \Y \times \U)$ with $X-U-Y$,
$R_X \geq 0$, $R_Y \geq 0$, $\delta > 0$, and $\textbf{u} \in
T^n_{P_U}$. Let
$\C^*_X=\{\textbf{x}_1,\textbf{x}_2,...,\textbf{x}_{M^*_X}\}$ and
$\C^*_Y=\{\textbf{y}_1,\textbf{y}_2,...,\textbf{y}_{M^*_Y}\}$ be the
collections of codewords whose existence is asserted in
lemma~\ref{packing2}. Consider the multiuser code
\begin{equation}
\C = \{ (\textbf{x}_i, \textbf{y}_j,D_{ij}) : i=1,...M^*_X,
j=1,...M^*_Y \}
\end{equation}
where the $D_{ij}$ are $\alpha$-decoding sets for $\textbf{u}$.
Taking into account the given $\textbf{u}$, the $\alpha$-decoding
yields the decoding sets
\begin{eqnarray}
D_{ij}= \{\textbf{z}: \alpha(\textbf{u},\textbf{x}_i,\textbf{y}_j,
\textbf{z}) \leq \alpha(\textbf{u},\textbf{x}_k,\textbf{y}_l,
\textbf{z}) \text{ for all} (k,l) \neq (i,j)\}\nonumber
\end{eqnarray}
The average probability of this multiuser code can be written as
\begin{align}
e(C,W) &\triangleq \frac{1}{M^*_X M^*_Y} \sum_{\substack{i,j}} W^n(D^c_{ij}|\textbf{x}_i,\textbf{y}_j)\nonumber\\
&= \frac{1}{M^*_X M^*_Y} \sum_{\substack{i,j}} W^n(\bigcup_{\substack{k \neq i}}D_{kj}|\textbf{x}_i,\textbf{y}_j)\nonumber\\
& +\frac{1}{M^*_X M^*_Y} \sum_{\substack{i,j}} W^n(\bigcup_{\substack{l \neq j}}D_{il}|\textbf{x}_i,\textbf{y}_j)\nonumber\\
& +\frac{1}{M^*_X M^*_Y} \sum_{\substack{i,j}}
W^n(\bigcup_{\substack{k \neq i\\ l \neq
j}}D_{kl}|\textbf{x}_i,\textbf{y}_j) \label{123}
\end{align}
The first term on the right side can be written as
\begin{eqnarray}
 \frac{1}{M^*_X M^*_Y} \sum_{\substack{i,j}} W^n\Big(\{\textbf{z}:\alpha(\textbf{u},\textbf{x}_i,\textbf{y}_j, \textbf{z}) > \alpha(\textbf{u},\textbf{x}_k,\textbf{y}_j, \textbf{z}),\nonumber\\ \text{for some } k  \neq i \}|\textbf{u}, \textbf{x}_i,\textbf{y}_j \Big)\nonumber\\
= \sum_{\substack{V_{UXY\tilde{X}Z} \in \V_X}} 2^{-n[D(V_{Z|XYU} || W|V_{XYU}) + H_V(Z|XYU)]}\nonumber\\
.\Big[ \frac{1}{M^*_X M^*_Y} \sum_{\substack{i,j}}
1_{T_{V_{UXY}}}(\textbf{u}, \textbf{x}_i,
\textbf{y}_j)\;\;\;\;\;\;\;\;\;\;\;\;\;\;\;\;\;\;\;\;\;\;\;\;\;\;\;\;\;\;\;\nonumber\\.\big|\{\textbf{z}:(\textbf{u},\textbf{x}_i,\textbf{y}_j,
\textbf{x}_k, \textbf{z}) \in T_{V_{UXY\tilde{X}Z}} \text{for some }
k \neq i\}\big|\Big]\label{form1-1}
\end{eqnarray}
The second term in~\eqref{form1-1} can be upper bounded by
\begin{eqnarray}
\frac{1}{M^*_X M^*_Y} \sum_{\substack{i,j}}
1_{T_{V_{UXY}}}(\textbf{u}, \textbf{x}_i,
\textbf{y}_j)\;\;\;\;\;\;\;\;\;\;\;\;\;\;\;\;\;\;\;\;\;\;\;\;\;\;\;\;\;\;\;\nonumber\\.\big|\{\textbf{z}:(\textbf{u},\textbf{x}_i,\textbf{y}_j,
\textbf{x}_k, \textbf{z}) \in T_{V_{UXY\tilde{X}Z}} \text{for some }
k \neq i\}\big|\nonumber\\
\leq \frac{1}{M^*_X M^*_Y} \sum_{\substack{i,j}} \sum_{k \neq i}
1_{T_{V_{UXY\tilde{X}}}}(\textbf{u}, \textbf{x}_i, \textbf{y}_j,
\textbf{x}_k) \;\;\;\;\;\;\;\;\;\;\;\;\;\;\; \nonumber \\
.\big|\{\textbf{z}:\textbf{z} \in
T_{V_{Z|UXY\tilde{X}}}(\textbf{u},\textbf{x}_i,\textbf{y}_j,
\textbf{x}_k\}\big|\label{final-1-1}
\end{eqnarray}
By properties of the codewords, mentioned in packing lemma, we can
bound the right side of~\eqref{final-1-1} by
\begin{align}
&\leq \exp{ \{-n\big[F_X(V)-H_V(Z|UXY\tilde{X})\big]\}}
\label{form1-1-gen}
\end{align}
By simple calculation, the exponent in~\eqref{form1-1-gen} can be
rewritten as
\begin{eqnarray}
I_V(X \wedge Y|U)+I_V(\tilde{X} \wedge Y|U)+I_V(\tilde{X} \wedge
XZ|UY)\nonumber\\-H_V(Z|UXY)-R_X
\end{eqnarray}
By using the fact that On the other hand, by the property of the
codebook, the following bound for the second term, on the right side
of~\eqref{form1-1}, can be obtained
\begin{align}
& \frac{1}{M^*_X M^*_Y} \sum_{\substack{i,j}}
1_{T_{V_{UXY}}}(\textbf{u}, \textbf{x}_i,
\textbf{y}_j)\;\;\;\;\;\;\;\;\;\;\;\;\;\;\;\;\;\;\;\;\;\;\;\;\;\;\;\;\;\;\;\nonumber\\&.\big|\{\textbf{z}:(\textbf{u},\textbf{x}_i,\textbf{y}_j,
\textbf{x}_k, \textbf{z}) \in T_{V_{UXY\tilde{X}Z}} \text{for some }
k \neq i\}\big|\nonumber\\
&\leq \exp{\big(-n[I_V(X \wedge Y|U)- H_V(Z|UXY) - 3
\delta]\big)}\label{form1-2}
\end{align}
By combining the exponents of~\eqref{form1-1-gen}
and~\eqref{form1-2}, the right side of~\eqref{form1-1} can be
bounded by
\begin{eqnarray}
\leq 2^{-nE_X(R_X,R_Y,W,P_{XYU})}\label{X-bnd}
\end{eqnarray}
where $E_X(R_X,R_Y,W,P_{XYU})$ is defined in~\eqref{Ex-def}.
Similarly, by using a similar argument for the second term on the
right side of \eqref{123}, we can show that
\begin{eqnarray}
\frac{1}{M^*_X M^*_Y} \sum_{\substack{i,j}} W^n(\bigcup_{\substack{l
\neq j}}D_{il}|\textbf{x}_i,\textbf{y}_j) \leq
2^{-nE_Y(R_X,R_Y,W,P_{XYU})} \label{Y-bnd}
\end{eqnarray}
where $E_Y(R_X,R_Y,W,P_{XYU})$ is defined in~\eqref{Ey-def}. Now,
consider the third term on the right side of ~\eqref{123}. It can be
written as
\begin{eqnarray}
\frac{1}{M^*_X M^*_Y} \sum_{\substack{i,j}} W^n\Big(\{\textbf{z}:\alpha(\textbf{u},\textbf{x}_i,\textbf{y}_j, \textbf{z}) > \alpha(\textbf{u},\textbf{x}_k,\textbf{y}_l, \textbf{z}),\nonumber\\ \text{for some } (k,l)  \neq (i,j) \}|\textbf{u}, \textbf{x}_i,\textbf{y}_j \Big)\nonumber\\
= \sum_{\substack{V_{UXY\tilde{X}\tilde{Y}Z} \in \V_{XY}}} 2^{-n[D(V_{Z|XYU} || W|V_{XYU}) + H_V(Z|XYU)]}\nonumber\\
.\Big[ \frac{1}{M^*_X M^*_Y} \sum_{\substack{i,j}}
1_{T_{V_{UXY}}}(\textbf{u}, \textbf{x}_i,
\textbf{y}_j)\;\;\;\;\;\;\;\;\;\;\;\;\;\;\;\;\;\;\;\;\;\;\;\;\;\;\;\;\;\;\;\nonumber\\.\big|\{\textbf{z}:(\textbf{u},\textbf{x}_i,\textbf{y}_j,
\textbf{x}_k, \textbf{y}_l, \textbf{z}) \in
T_{V_{UXY\tilde{X}\tilde{Y}Z}} \nonumber \\ \text{for some }(k,l)
\neq (i,j)\}\big|\Big]\label{form3-1}
\end{eqnarray}
The second term in~\eqref{form3-1} can be upper bounded by
\begin{eqnarray}
\leq \frac{1}{M^*_X M^*_Y} \sum_{\substack{i,j}} \sum_{\substack{k
\neq i\\l \neq j}} 1_{T_{V_{UXY\tilde{X}\tilde{Y}}}}(\textbf{u},
\textbf{x}_i, \textbf{y}_j,
\textbf{x}_k,\textbf{y}_l) \;\;\;\;\;\;\;\;\;\;\;\;\;\;\; \nonumber \\
.\big|\{\textbf{z}:\textbf{z} \in
T_{V_{Z|UXY\tilde{X}\tilde{Y}}}(\textbf{u},\textbf{x}_i,\textbf{y}_j,
\textbf{x}_k,\textbf{y}_l)\}\big|\label{final-3-1}
\end{eqnarray}
The second term is actually the cardinality of
$T_{V_{Z|UXY\tilde{X}\tilde{Y}}}(\textbf{u},\textbf{x}_i,\textbf{y}_j,
\textbf{x}_k,\textbf{y}_l)$, which is equal to $\exp{\{n
H_V(Z|UXY\tilde{X}\tilde{Y})\}}$. By the properties of the
codewords, the first term in~\eqref{form3-1} can be upper bounded by
$\exp{\{-n[F_{XY}(V)]\}}$. Therefore~\eqref{final-3-1} can be
bounded by
\begin{align}
&\leq \exp{ \{-n\big[F_{XY}(V)-H_V(Z|UXY\tilde{X}\tilde{Y})\big]\}}
\label{form3-1-gen}
\end{align}
By simple calculation, the exponent in~\eqref{form3-1-gen} can be
rewritten as
\begin{eqnarray}
I_V(X \wedge Y|U)+I_V(\tilde{X} \wedge
\tilde{Y}|U)+I_V(\tilde{X}\tilde{Y} \wedge
XYZ|U)\nonumber\\-H_V(Z|UXY)-R_X-R_Y
\end{eqnarray}
By using the properties of codewords, the following bound for the
second term on the right right side of~\eqref{form3-1} can be
obtained as follows
\begin{align}
& \frac{1}{M^*_X M^*_Y} \sum_{\substack{i,j}}
1_{T_{V_{UXY}}}(\textbf{u}, \textbf{x}_i,
\textbf{y}_j)\;\;\;\;\;\;\;\;\;\;\;\;\;\;\;\;\;\;\;\;\;\;\;\;\;\;\;\;\;\;\;\nonumber\\&\big|\{\textbf{z}:(\textbf{u},\textbf{x}_i,\textbf{y}_j,
\textbf{x}_k, \textbf{y}_l, \textbf{z}) \in
T_{V_{UXY\tilde{X}\tilde{Y}Z}} \text{for some }
(k,l) \neq (i,j)\}\big|\nonumber\\
&\leq \exp{\big(-n[I_V(X \wedge Y|U)- H_V(Z|UXY) - 3
\delta]\big)}\label{form3-2}
\end{align}
By combining the exponents of~\eqref{form3-1-gen}
and~\eqref{form3-2}, the right side of~\eqref{form3-1} can be
bounded by
\begin{eqnarray}
\leq 2^{-nE_{XY}(R_X,R_Y,W,P_{XYU})}\label{XY-bnd}
\end{eqnarray}
where $E_{XY}(R_X,R_Y,W,P_{XYU})$ is defined in~\eqref{Exy-def}.
Now, it follows from~\eqref{X-bnd},~\eqref{Y-bnd},
and~\eqref{XY-bnd}, that the average probability of the given code
is upper bounded by
\begin{eqnarray}
e(C,W) \leq 2^{-n[E_{ex}(R_X,R_Y,W,P_{XYU})- \delta]}
\end{eqnarray}
where $E_{ex}(R_X,R_Y,W,P_{XYU})$ is defined
in~\eqref{Expuragted-def}.
\end{proof}
In the following, we prove that the random coding bound in
theorem~\ref{randomcodingthm} will result in a tighter bound in
comparison to the best known random coding bound, found
in~\cite{Liu-RandomCoding}. For this purpose, let us use the minimum
equivocation decoding rule.
\begin{definition}
Given $\textbf{u}$, for a multiuser code
\begin{equation*}
\C = \{ (\textbf{x}_i, \textbf{y}_j,D_{ij}) : i=1,...M^*_X,
j=1,...M^*_Y \}
\end{equation*}
we say that the $D_{ij}$ are minimum equivocation decoding sets for
$\textbf{u}$ if $\textbf{z} \in D_{ij}$ implies
\begin{equation*}
H(\textbf{x}_i \textbf{y}_j|\textbf{z} \textbf{u}) = \min_{k,l}
H(\textbf{x}_k \textbf{y}_l|\textbf{z} \textbf{u}).
\end{equation*}
It can be easily observed that these sets are equivalent to
$\alpha$-decoding sets, where $\alpha(\textbf{u}, \textbf{x},
\textbf{y}, \textbf{z})$ is defined as
\begin{equation}
\alpha(V_{UXYZ}) \triangleq H_V(XY|ZU).
\end{equation}
Here, $V_{UXYZ}$ is the joint empirical distribution of
$(\textbf{u}, \textbf{x}, \textbf{y}, \textbf{z})$.
\end{definition}
\begin{theorem}
 For every finite set
$\U$, $\P_{XYU} \in \P(\U)$ , $R_X \geq 0$, $R_Y \geq 0$, and $W:\X
\times \Y \rightarrow \Z$,
\begin{eqnarray}
E_{\beta}(R_X,R_Y,W,P_{XYU}) \geq
E^L_{r\beta}(R_X,R_Y,W,P_{XYU})\nonumber\\
\beta=X,Y,XY
\end{eqnarray}
Hence
\begin{equation}
E_{ex}(R_X,R_Y,W,P_{XYU}) \geq E^L_r(R_X,R_Y,W,P_{XYU})
\end{equation}
for all $P_{XYU} \in \P(\U)$ satisfying $X-U-Y$. Here, $E^L_{r}$ is
the random coding exponent of~\cite{Liu-RandomCoding}.
$E^L_{r\beta}$ are also defined in~\cite{Liu-RandomCoding}.
\end{theorem}
\begin{proof}
For any $\substack{V_{UXY\tilde{X}Z}} \in \V_X$,
\begin{eqnarray}
H_V(XY|ZU) \geq H_V(\tilde{X}Y|ZU),\label{equivocation-rule}
\end{eqnarray}
therefore, by subtracting $H_{V}(Y|ZU)$ form both sides of
\eqref{equivocation-rule}, we can conclude that
\begin{eqnarray*}
H_V(X|U)- I_V(X \wedge YZ|U) \geq H_V(\tilde{X}|U)- I_V(\tilde{X}
\wedge YZ|U), \label{Gen-Compare-Liu-X1}
\end{eqnarray*}
Since $V_{XU}=V_{\tilde{X}U}=P_{XU}$, the last inequality is
equivalent to
\begin{equation*} I_V(X \wedge YZ|U) \leq I_V(\tilde{X}
\wedge YZ|U)
\end{equation*}
Since $I_V(\tilde{X} \wedge XZ|YU) + I(\tilde{X} \wedge Y|U) \geq
I_V(\tilde{X} \wedge YZ|U)$, it can be seen that for any
$\substack{V_{UXY\tilde{X}Z}} \in \V_X$
\begin{equation*}
I_V(\tilde{X} \wedge XZ|YU) + I(\tilde{X} \wedge Y|U) \geq I_V(X
\wedge YZ|U)
\end{equation*}
Moreover, since
\begin{eqnarray}
\V_X \subset \{V_{UXY\tilde{X}Z}: V_{UXYZ} \in
\V(P_{UXY})\nonumber\\
I(X \wedge Y|U) \leq R_X+ 3 \delta \}
\end{eqnarray}
 it can be easily concluded that
\begin{equation*}
E_{X}(R_X,R_Y,W,P_{XYU}) \geq E^L_{rX}(R_X,R_Y,W,P_{XYU}).
\end{equation*}
Similarly, for any $V_{UXY\tilde{Y}Z} \in \V_Y$,
\begin{eqnarray*}
H_V(XY|ZU) \geq H_V(X\tilde{Y}|ZU).
\end{eqnarray*}
By using the fact that, $V_{YU}=V_{\tilde{Y}U}=P_{YU}$, it can be
concluded that
\begin{equation*}
I_V(\tilde{Y} \wedge YZ|XU) + I(X \wedge \tilde{Y}|U) \geq I_V(Y
\wedge XZ|U).
\end{equation*}
Since
\begin{eqnarray}
\V_Y \subset \{V_{UXY\tilde{Y}Z}: V_{UXYZ} \in
\V(P_{UXY})\nonumber\\
I(X \wedge Y|U) \leq R_X+ 3 \delta \}
\end{eqnarray}
we conclude that
\begin{equation*}
E_{Y}(R_X,R_Y,W,P_{XYU}) \geq E^L_{rY}(R_X,R_Y,W,P_{XYU}).
\end{equation*}
Similarly, we can conclude that, for any $V_{UXY\tilde{X}\tilde{Y}Z}
\in \V_{XY}$,
\begin{equation*}
I_V(\tilde{X}\tilde{Y} \wedge XYZ|U) + I(\tilde{X} \wedge
\tilde{Y}|U) \geq I_V(XY \wedge Z|U) + I(X \wedge Y|U).
\end{equation*}
Since
\begin{eqnarray}
\V_{XY} \subset \{V_{UXY\tilde{X}\tilde{Y}Z}: V_{UXYZ} \in
\V(P_{UXY})\nonumber\\
I(X \wedge Y|U) \leq R_X+ 3 \delta \},
\end{eqnarray}
it can be concluded that
\begin{equation*}
E_{XY}(R_X,R_Y,W,P_{XYU}) \geq E^L_{rXY}(R_X,R_Y,W,P_{XYU}).
\end{equation*}
\end{proof}
The last theorem shows that $E_{ex}(R_X,R_Y,W,P_{XYU})$ is at least
as large as the Liu, Hughes~\cite{Liu-RandomCoding} exponent. In the
following, we show that at low rate pairs, we may have a strictly
better result. To illustrate this, let us focus on the case where
both codebooks have rate zero, $R_X=R_Y=0$. For small $\delta$, any
$V_{UXY\tilde{X}Z} \in \V_X$ will satisfy the following
relationships
\begin{equation}
X-U-Y, \;\;\;  \tilde{X}-U-Y, \;\;\; \tilde{X}-UY-X
\end{equation}
Therefore, any $V_{UXY\tilde{X}Z} \in \V_X$ can be written as
\begin{equation}
V_{Z|UXY\tilde{X}}P_{X|U}P_{Y|U}P_{X|U}P_{U}.
\end{equation}
Similarly, any $V_{UXY\tilde{Y}Z} \in \V_Y$ can be written as
\begin{equation}
V_{Z|UXY\tilde{Y}}P_{X|U}P_{Y|U}P_{Y|U}P_{U},
\end{equation}
and any $V_{UXY\tilde{X}\tilde{Y}Z} \in \V_{XY}$ can be written as
\begin{equation}
V_{Z|UXY\tilde{X}\tilde{Y}}P_{X|U}P_{Y|U}P_{X|U}P_{Y|U}P_{U}.
\end{equation}

For a moment, let us consider the point to point data communication.
By using only a random coding argument, and without any expurgation,
one can prove the following result.
\begin{lemma}
For every $R > 0$, $\delta \geq 0$ and every type of $P \in P_n(\X)$
satisfying $H(P) \geq R$, there exist $M \geq 2^{n(R-\delta)}$
sequences in $T_P$ such that for every $P_{X\tilde{X}} \in \P(\X
\times \X)$,
\begin{equation}
\frac{1}{M} \sum_{i=1}^{M} \sum_{k \neq i} 1_{T_{P_{X\tilde{X}}}}
(\textbf{x}_i,\textbf{x}_k) \leq 2^{n(R-I(X \wedge \tilde{X}))}
\label{pp-packing}
\end{equation}
provided that $n \geq n_0(|\X|,|\Y|,\delta)$.
\end{lemma}
Now, let us multiply both sides of~\eqref{pp-packing} by $M$. It can
be shown that for every $1 \leq i \leq M$,
\begin{equation}
\sum_{k \neq i} 1_{T_{P_{X\tilde{X}}}} (\textbf{x}_i,\textbf{x}_k)
\leq 2^{n(2R-I(X \wedge \tilde{X}))}
\end{equation}
By using these sequences as our set of codewords, and using
$\alpha$-decoding, we will end up with a result very similar to
\cite{Csiszar-Graph}. The only difference is that our minimization
would be taken over all distributions satisfying $I(X \wedge
\tilde{X}) \leq 2R$, instead of $I(X \wedge \tilde{X}) \leq R$.
Using the appropriate decoding rule, this bound would be exactly the
same as the typical random coding bound that Barg and Forney found
in \cite{Barg-RandomCode}. As we can see, in point to point
communications, even without doing any expurgation, we ended up with
a strictly better bound in comparison to the usual random coding
bound. Needless to say that if we eliminate half of the codewords
in~\eqref{pp-packing}, the result would be equal to the expurgated
bound~\cite{Csiszar-Graph}.

\bibliographystyle{plain}
\bibliography{ali}

\begin{thebibliography}{10}

\bibitem{Ahlswede71}
R.~Ahlswede.
\newblock Multi-way communication channels.
\newblock In {\em Proc.~International Symposium on Information Theory}, 1971.

\bibitem{Barg-RandomCode}
A.~Barg and D.~Forney.
\newblock Random codes: Minimum distances and error exponents.
\newblock {\em IEEE Trans.~Information Theory}, 48(9):2568--2573, Sept. 2002.

\bibitem{Csiszar-Graph}
I.~Csiszar and J.~Korner.
\newblock Graph decomposition: A new key to coding theorems.
\newblock {\em IEEE Trans.~Information Theory}, 1:5--12, Jan. 1981.

\bibitem{Csiszarbook}
I.~Csiszar and J.~Korner.
\newblock {\em Information theory: Coding theorems for Discrete memoryless
  Systems.}
\newblock 1981.

\bibitem{Dyachkov}
A.~G. Dyachkov.
\newblock Random constant composition codes for multiple-access channels.
\newblock {\em Probl.~of Control and Inform.~Theory}, pages 357--369, 1984.

\bibitem{Gallager-Tightness}
R.~Gallager.
\newblock The random coding bound is tight for the average code.
\newblock {\em IEEE Trans.~Information Theory}, 23(2):244--246, Mar. 1973.

\bibitem{Gallager-Multiaccess}
R.~Gallager.
\newblock A perspective on multi-access channels.
\newblock {\em IEEE Trans.~Information Theory}, 31(2):124--142, Mar. 1985.

\bibitem{FanoBook}
R.~G. Gallager.
\newblock {\em Transmission of Information: A Statistical Theory of Communica-
  tion.}
\newblock MIT Press, 1961.

\bibitem{GalBook}
R.~G. Gallager.
\newblock {\em Information theory and Reliable Communications}.
\newblock John Wiley \& Sons, New York, 1968.

\bibitem{Haroutunian}
E.~A. Haroutunian.
\newblock Lower bound for the error probability of multiple-access channels.
\newblock {\em Problemy Peredachi Informatsii}, 11:23--36, June 1975.

\bibitem{Liao}
H.~Liao.
\newblock A coding theorem for multiple-access communications.
\newblock In {\em Proc.~International Symposium on Information Theory}.

\bibitem{Liu-RandomCoding}
Y.~Liu and B.~L. Hughes.
\newblock A new universal random coding bound for the multiple-access channels.
\newblock {\em IEEE Trans.~Information Theory}, 42(2):376--386, Mar. 1996.

\bibitem{nazari08}
Ali Nazari, S.~Sandeep Pradhan, and Achilleas Anastasopoulous.
\newblock A new sphere-packing bound for maximal error exponent for
  multiple-access channels.
\newblock In {\em Proc.~International Symposium on Information Theory}, 2008.
\newblock Online: http://arxiv.org/abs/0803.3645.

\bibitem{Pokorney}
J.~Pokorney and H.~S. Wallmeier.
\newblock Random coding bounds and codes produced by permutations for the
  multiple-access channels.
\newblock {\em IEEE Trans.~Information Theory}, 31(6):741--750, Nov. 1985.

\bibitem{SlWo73}
D.~Slepian and J.~K. Wolf.
\newblock A coding theorem for multiple access channels with correlated
  sources.
\newblock {\em bell Syst. tech. J.}, 52:1037--1076, 1973.

\end{thebibliography}

\end{document}